\documentclass[runningheads]{svmult}
\usepackage{makeidx}   
\usepackage{graphicx}  
\usepackage{subeqnar}  
\usepackage{multicol}  
\usepackage{physprbb}  
\makeindex             

\def\lesssim{\mathrel{\hbox{\rlap{\hbox{\lower4pt\hbox{$\sim$}}}\hbox{$<$}}}}
\def\gtrsim{\mathrel{\hbox{\rlap{\hbox{\lower4pt\hbox{$\sim$}}}\hbox{$>$}}}}

\begin{document}
\title*{Theory of GRB Afterglow}
\titlerunning{Theory of GRB Afterglow}

\author{T. Piran \and J. Granot}
\authorrunning{Tsvi Piran and Jonathan Granot}
%
%
\institute{Racah Institute of Physics, Hebrew University, Jerusalem 91904, Israel}

\maketitle              

\vspace{-0.5cm}

\begin{abstract}
  \index{abstract}
The most interesting current open question in the theory of GRB
afterglow is the propagation of jetted afterglows during the
sideway expansion phase.  Recent numerical simulations show
hydrodynamic behavior that differs from the one suggested by
simple analytic models. Still, somewhat surprisingly, the
calculated light curves show a `jet break' at about the expected
time. These results suggest that the expected rate of orphan
optical afterglows should be smaller than previously estimated.
\end{abstract}

\section{Introduction}
\label{intro}

Our understanding of GRBs has been revolutionized by the BeppoSAX
discovery of GRB afterglow. While GRBs last seconds or minutes the
afterglow lasts days, weeks months or even years. This makes afterglow
observations much richer. These observations provide us with
multi-wavelength and multi-timescales data. At the same time the
afterglow, which is a blast wave propagating into the surrounding
matter is a much simpler phenomena than the GRB and it is possible to
construct a simple theory that can be compared directly with the
observations.

In this short review we describe the theory of GRB afterglow. We begin
with the simplest idealized model and continue with various levels of
complications. The final level is full numerical simulations. We
present preliminary results of such simulations and compare them with
analytic models.  At present there is no simple analytic explanation
for the features seen in the numerical results.

\section{Spherical Hydrodynamics}

The theory of relativistic blast waves has been worked out in a
classical paper by Blandford \& McKee (BM) already in 1976
\cite{BM76}. The BM model is a self-similar spherical solution
describing an adiabatic ultra relativistic blast wave in the
limit $\Gamma \gg 1$.  The basic solution is a blast wave
propagating into a constant density medium. However, Blandford
and McKee also describe in the same paper a generalization for
varying ambient mass density, $\rho =A R^{-k}$, $R$ being the
distance from the center. The latter case would be particularly
relevant for $k=2$, as expected in the case of wind from a
progenitor, prior to the GRB explosion.

The BM solution describes a narrow shell of width $\sim
R/\Gamma^2$, in which the shocked material is concentrated, where
$\Gamma$ is the typical Lorentz factor. The conditions in this
shell can be approximated if we assume that the shell is
homogeneous.  Then the adiabatic energy conservation yields:
\begin{equation}
E =  {\Omega\over 3-k} A R^{3-k} \Gamma^2 c^2 \ , \label{ad}
\end{equation}
where $E$ is the energy of the blast wave and $\Omega$ is the solid
angle of the afterglow. For a full sphere $\Omega= 4\pi$, but it can
be smaller if the expansion is conical with an opening angle $\theta$:
$\Omega = 2 \pi \theta^2$ (assuming a double sided jet).

A natural length scale, $l=\left[(3-k)E/\Omega A c^2\right]^{1/(3-k)}$,
appears in equation \ref{ad}. For a spherical blast wave $\Omega$
does not change with time, and when the blast wave reaches $R=l$ it
collects ambient rest mass that equals its initial energy, the
Lorentz factor $\Gamma$ drops to 1 and the blast wave becomes
Newtonian. The BM solution is self-similar and assumes $\Gamma
\gg 1$.  Obviously, it breaks down when $R\sim l$. We therefore
expect that a Relativistic-Newtonian transition should take place
around  $t_{\rm
  NR}=l/c \approx 1.2 \, {\rm yr} (E_{\rm iso,52}/n_1)^{1/3}$, where
the scaling is for $k=0$, $E_{52}$ is the isotropic equivalent
energy, $E_{\rm iso}=4\pi E/\Omega$, in units of $10^{52} {\rm
ergs}$ and $n_1$ is the external density in ${\rm cm}^{-3}$. After
this transition the solution will turn into the Newtonian
Sedov-Tailor solution. Clearly this produces an achromatic break
in the light curve.

The adiabatic approximation is valid for most of the duration of
the fireball. However, during the first hour or so (or even for
the first day, for $k = 2$), the system could be radiative
(provided that $\epsilon_e \approx 1$). During a radiative phase
the evolution can be approximated as:
\begin{equation}
E =  {\Omega\over 3-k} A R^{3-k} \Gamma \Gamma_0 c^2 \ , \label{rad}
\end{equation}
where $\Gamma_0$ is the initial Lorentz factor. Cohen, Piran \& Sari
\cite{CPS98} derived an analytic self-similar solution describing this
phase. Cohen \& Piran \cite{CP99} describe a solution for the case when
energy is continuously added to the blast wave by the central engine,
even during the afterglow phase. A self-similar solution arises if the
additional energy deposition behaves like a power law.  This would
arise naturally in some models, e.g. in the pulsar like model \cite{Usov94}.

\section{Spherical Afterglow Models}

A good model for the observed emission from spherical blast waves
can be obtained by adding synchrotron radiation to these
hydrodynamic models. Sari, Piran  \& Narayan \cite{SPN98} used the
simple adiabatic scaling (\ref{ad}) together with synchrotron
radiation model and the relation between the observer time $t$, and
$R$:
\begin{equation}
t =R / C_1  c \Gamma^{2} \ , \label{tobs}
\end{equation}
where $C_1$ is a constant that may vary from $2$ to $16$
\cite{Sari97}.

Assuming a powerlaw energy distribution of the shocked
relativistic electrons: $N(E_e)\propto E_e^{-p}$, and that the
electrons and the magnetic field energy densities are
$\epsilon_e$ and $\epsilon_B$ times the total energy density,
Sari, Piran \& Narayan \cite{SPN98} estimate the observed
emission as a series of  power law segments (PLSs), where
\begin{equation}
 F_\nu \propto  t^{-\alpha} \nu^{-\beta} \ ,
\end{equation}
that are separated by break frequencies, across which the
exponents of these power laws change: the cooling frequency,
$\nu_c$, the typical synchrotron frequency $\nu_m$ and the self
absorption frequency $\nu_{sa}$. The analytic calculations were
done for a homogeneous shell and for emission from a single
representative point. At a specific frequency one will observe a
break in the light curve when one of these break frequencies
passes the observed frequency. An intriguing feature of this
model is that for a given PLS, say for emission above the cooling
frequency, there is a unique relation between $\alpha$, $\beta$
and $p$. The power law index $p$ is expected to be a universal
quantity as it depends on the, presumably common, acceleration
processes and it is expected to be between 2 and 2.5 \cite{SP97}.
The consistency of those observed parameters could be a simple
check of the theory.

The simple solution, that is based on a homogeneous shell
approximation, can modified by using the full BM solution and
integrating over the entire volume of shocked fluid \cite{GPS99}. Such
an integration can be done only numerically. It yields a smoother
spectrum and light curve near the break frequencies, but the
asymptotic slopes, away from the break frequencies and the transition
times, remain the same as in the simpler theory.

Chevalier \& Lee \cite{CheLi99} estimated the emission from a
blast wave propagating into a wind profile $n(R) \propto
R^{-2}$.  They use equation (\ref{ad}) and calculate the
synchrotron emission from a single representative point. This
leads to different temporal scalings $\alpha$ of the PLSs, while
the spectral indices $\beta$ remain the same, since they are
independent of the hydrodynamic solution. This results in
different relations between $\alpha$, $\beta$ and $p$, providing
in principle a way to distinguish between different neighborhoods
of GRBs and between different progenitor models.

Another modification to the "standard" model arises from a
variation of the emission process. Sari \& Esin \cite{SariEsin01} considered
the influence of Inverse Compton on the observed spectrum. They
find that in some cases the additional cooling channel might have
a significant effect on the observed spectrum and light curves.

\section{Jets}

The afterglow theory becomes much more complicated if the
relativistic ejecta is not spherical. To model jetted afterglows
we consider relativistic matter ejected into a cone of opening
angle $\theta$. Initially, as long as $\Gamma \gg \theta^{-1}$
\cite{Pi94} the motion would be almost conical. There isn't
enough time, in the blast wave's rest frame, for the matter to be
affected by the non spherical geometry, and the blast wave will
behave as if it was a part of a sphere. When $\Gamma = C_2
\theta^{-1}$, namely at\footnote{The exact values of the uncertain
constants $C_2$ and $C_1$ are extremely important as they
determine the jet opening angle (and hence the total energy of
the GRB) from the observed breaks, interpreted as $t_{\rm jet}$, in
the afterglow light curves.}:
\begin{equation}
t_{\rm jet}  =   {1\over C_1}\left( l\over c \right )
\left({\theta\over C_2}\right)^{2(4-k)\over (3-k)} =  {1 \, {\rm
day} \over C_1 C_2^{8/3}} \left({E_{\rm iso,52}\over
n_1}\right)^{1/3} \left({\theta\over 0.1}\right)^{8/3} \ ,
\label{tjet}
\end{equation}
rapid sideway propagation begins. The last equality holds, of
course for $k=0$.

The sideways expansion continues with $\theta \sim \Gamma^{-1}$.
Plugging this relations in equation (\ref{ad}) we find that $R
\approx {\rm const}.$ This is obviously impossible. A more
detailed analysis \cite{Rhoads99,Pi00,KP00} reveals that
according to the simple one dimensional analytic models $\Gamma$
decreases exponentially with $R$ on a very short length
scale.\footnote{Note that the exponential behavior is obtained after
converting equation \ref{ad} to a differential equation and
integrating over it. Different approximations used in deriving
the differential equation lead to slightly different exponential
behavior, see \cite{Pi00}.}

The sideways expansion causes a change in the hydrodynamic
behavior and hence a break in the light curve. Additionally, when
$\Gamma \sim \theta^{-1}$ relativistic beaming of light will
become less effective. This would cause an extra spreading of the
emission (that was previously focused into a narrow angle
$\theta$ and is now focused into a larger cone of opening angle
$\Gamma^{-1}$). If the sideways expansion is at the speed of
light than both transitions would take place at the same time
\cite{SPH99}. If the sideways expansion is at the sound speed
then the beaming transition would take place first and only later
the hydrodynamic transition would occur \cite{PM99}.  This would
cause a slower and wider transition with two distinct breaks, the
first and steeper break when the edge of the jet becomes visible
and later a shallower break when sideways expansion becomes
important.

The  analytic or semi-analytic calculations of synchrotron
radiation from jetted afterglows
\cite{Rhoads99,SPH99,PM99,MSB00,KP00} have led to different
estimates of  the jet break time $t_{\rm jet}$ and of the
duration of the transition. Rhoads \cite{Rhoads99} calculated the
light curves assuming emission from one representative point, and
obtained a smooth 'jet break', extending $\sim 3-4$ decades in
time, after which $F_{\nu>\nu_m}\propto t^{-p}$. Sari Piran \&
Halpern \cite{SPH99} assume that the sideway expansion is at the
speed of light, and not at the speed of sound ($c/\sqrt{3}$) as
others assume, and find a  smaller value for $t_{\rm jet}$.
Panaitescu and M\'esz\'aros \cite{PM99} included the effects of
geometrical curvature and finite width of the emitting shell,
along with electron cooling, and obtained a relatively sharp
break, extending $\sim 1-2$ decades in time, in the optical light
curve. Moderski, Sikora and Bulik \cite{MSB00} used a slightly
different dynamical model, and a different formalism for the
evolution of the electron distribution, and obtained that the
change in the temporal index $\alpha$ ($F_{\nu}\propto
t^{-\alpha}$) across the break is smaller than in analytic
estimates ($\alpha=2$ after the break for $\nu>\nu_m$, $p=2.4$),
while the break extends over two decades in time.  Kumar and
Panaitescu \cite{KP00} find that for a homogeneous (or stellar
wind) environment there is a steepening of $\Delta\alpha\sim 0.7$
($0.4$) when the edge of the jet becomes visible, while the
steepening due to sideways expansion extends over 2 (4) decades
in time. They conclude that a jet running into a stellar wind
will not leave a prominent detectable signature in the light
curve.

\begin{figure}
\begin{center}
\includegraphics[width=4.97cm]{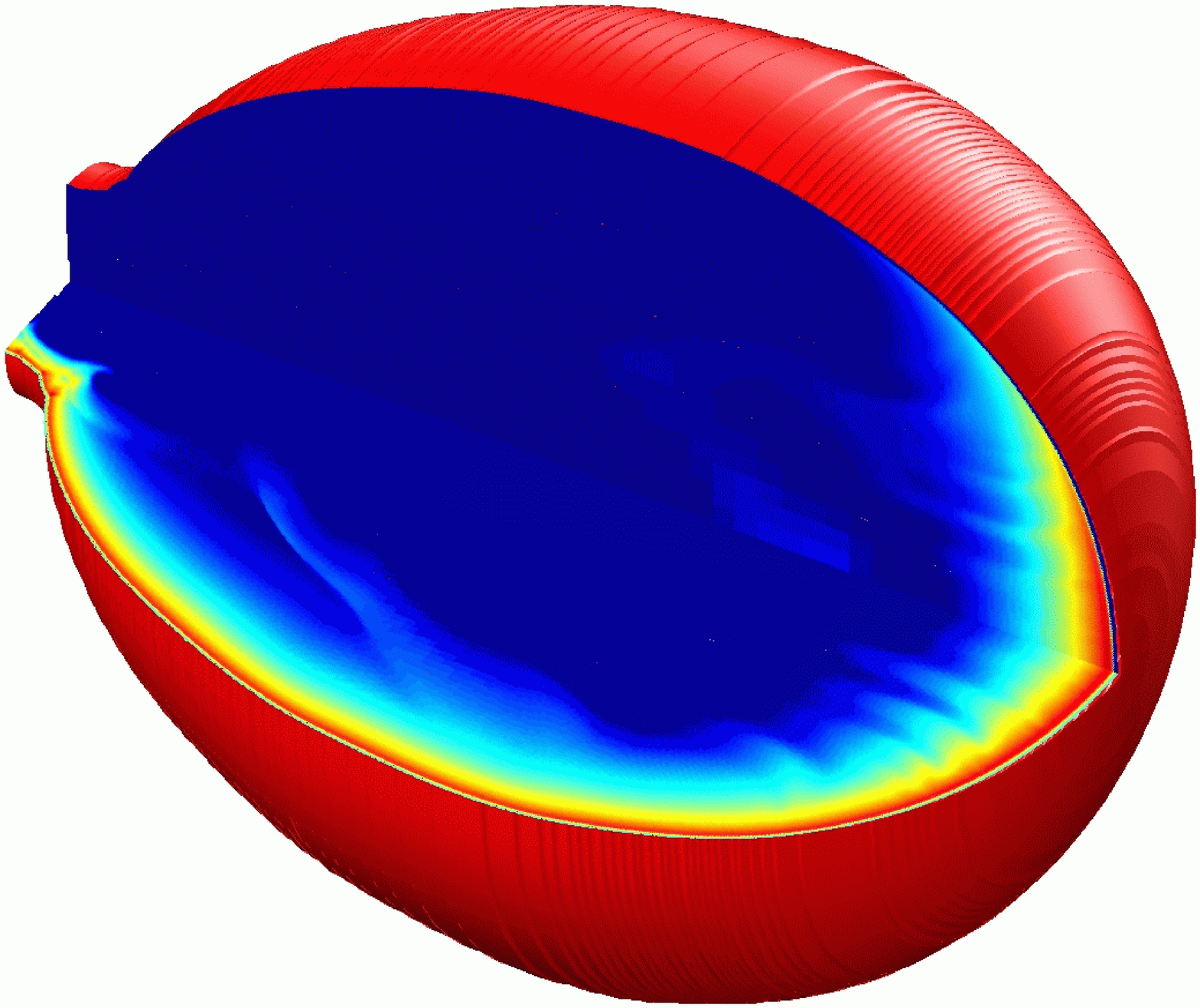}
\includegraphics[width=7.0cm]{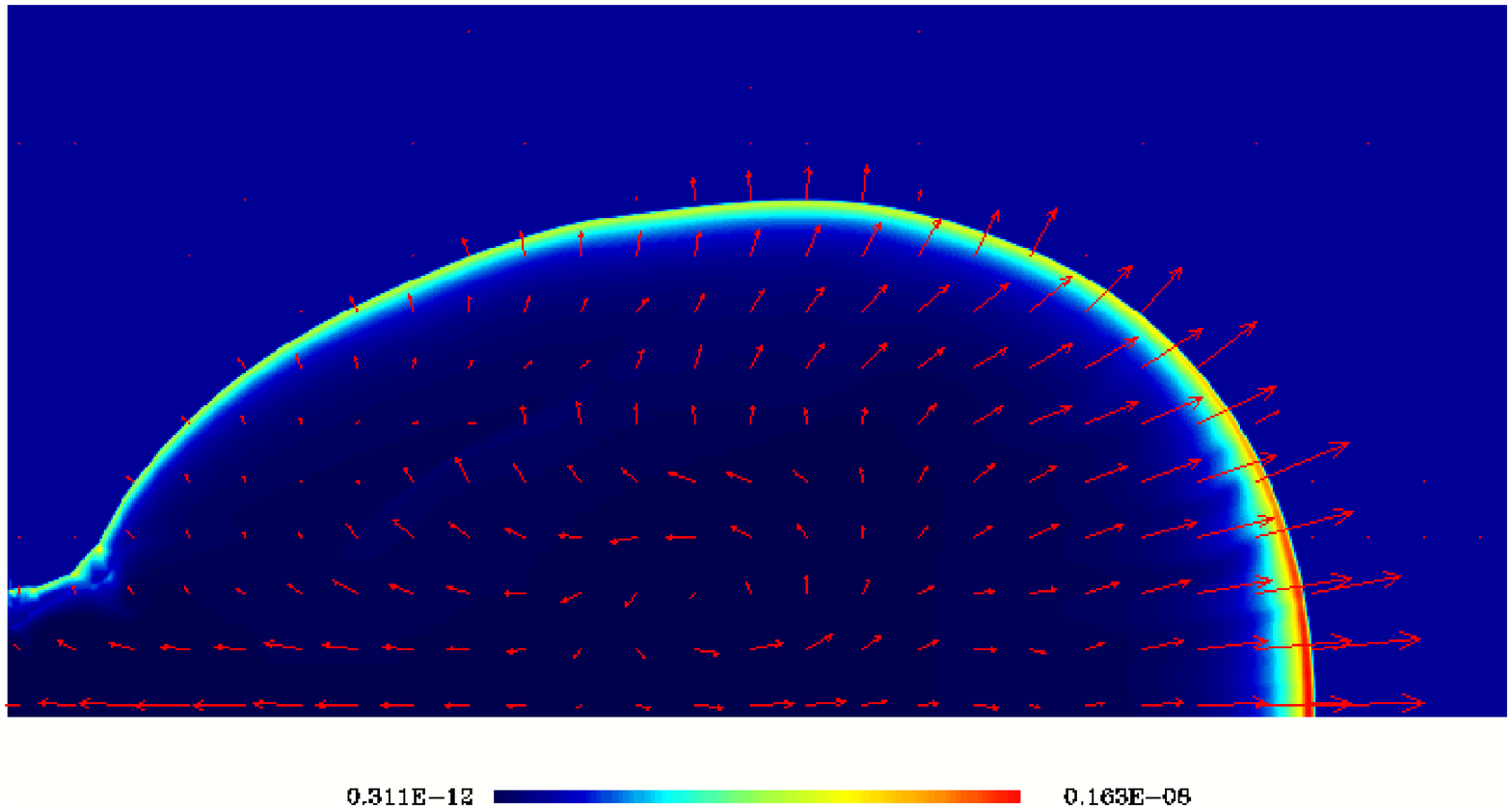}
\end{center}
\caption[]{A relativistic jet at the last time
  step of the simulation \cite{Granot01}.  ({\bf left}) A 3D view
  of the jet. The outer surface represents the shock front while the
  two inner faces show the proper number density ({\it lower face})
  and proper emissivity ({\it upper face}) in a logarithmic color
  scale. ({\bf right}) A 2D 'slice' along the jet axis, showing the
  velocity field on top of a linear color-map of the lab frame
  density.}
\label{3Djet}
\end{figure}

The different analytic or semi-analytic models have different
predictions  for the sharpness of the 'jet break', the change in
the temporal decay index $\alpha$ across the break and its
asymptotic value after the break, or even the very existence a
'jet break' \cite{HDL00}. All these models rely on some common
basic assumptions, which have a significant effect on the
dynamics of the jet: (i) the shocked matter is homogeneous (ii)
the shock front is spherical (within a finite opening angle) even
at $t>t_{\rm jet}$ (iii) the velocity vector is almost radial even
after the jet break.

\begin{figure}
\begin{center}
\includegraphics[width=3.97cm]{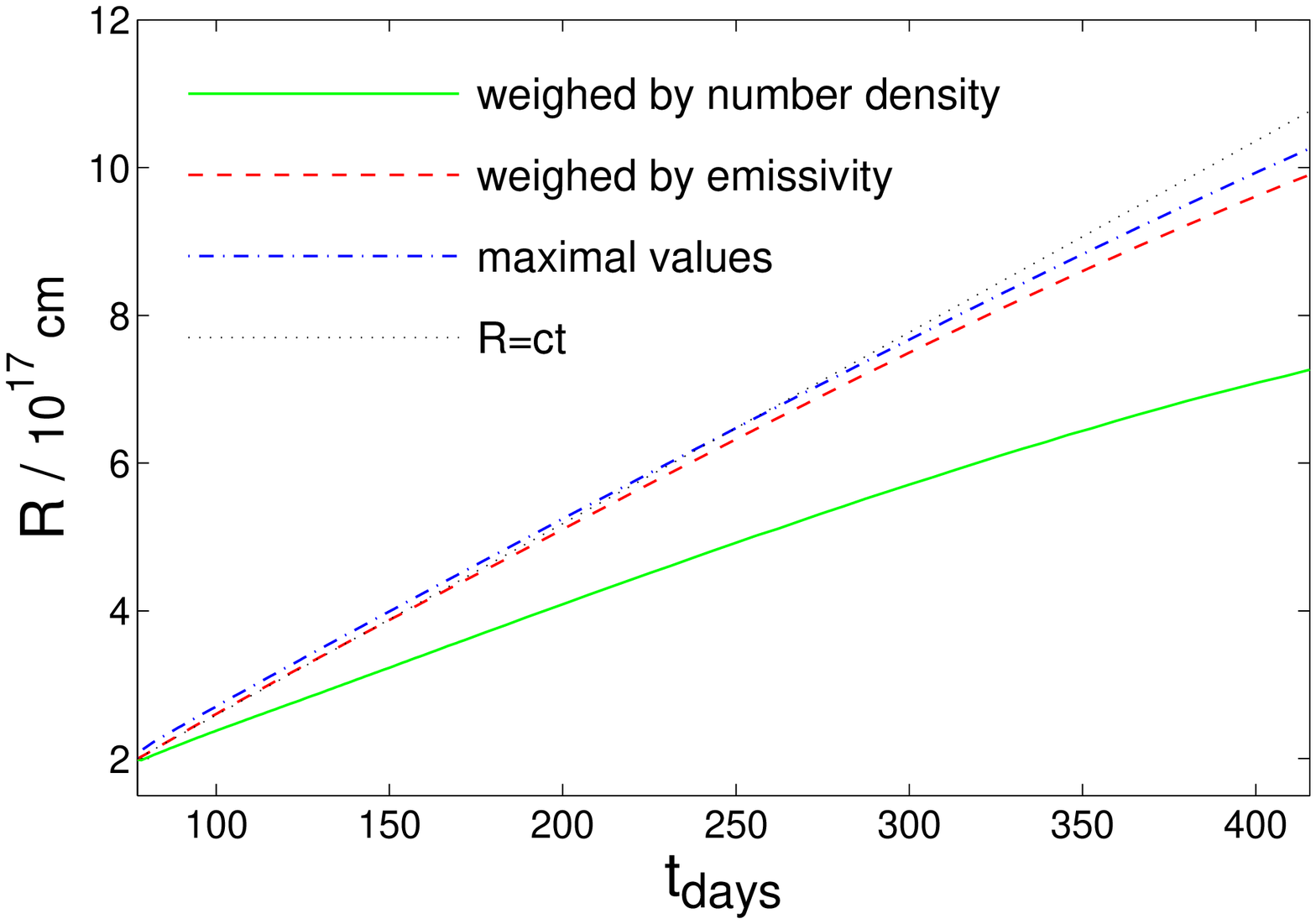}
\includegraphics[width=4.05cm]{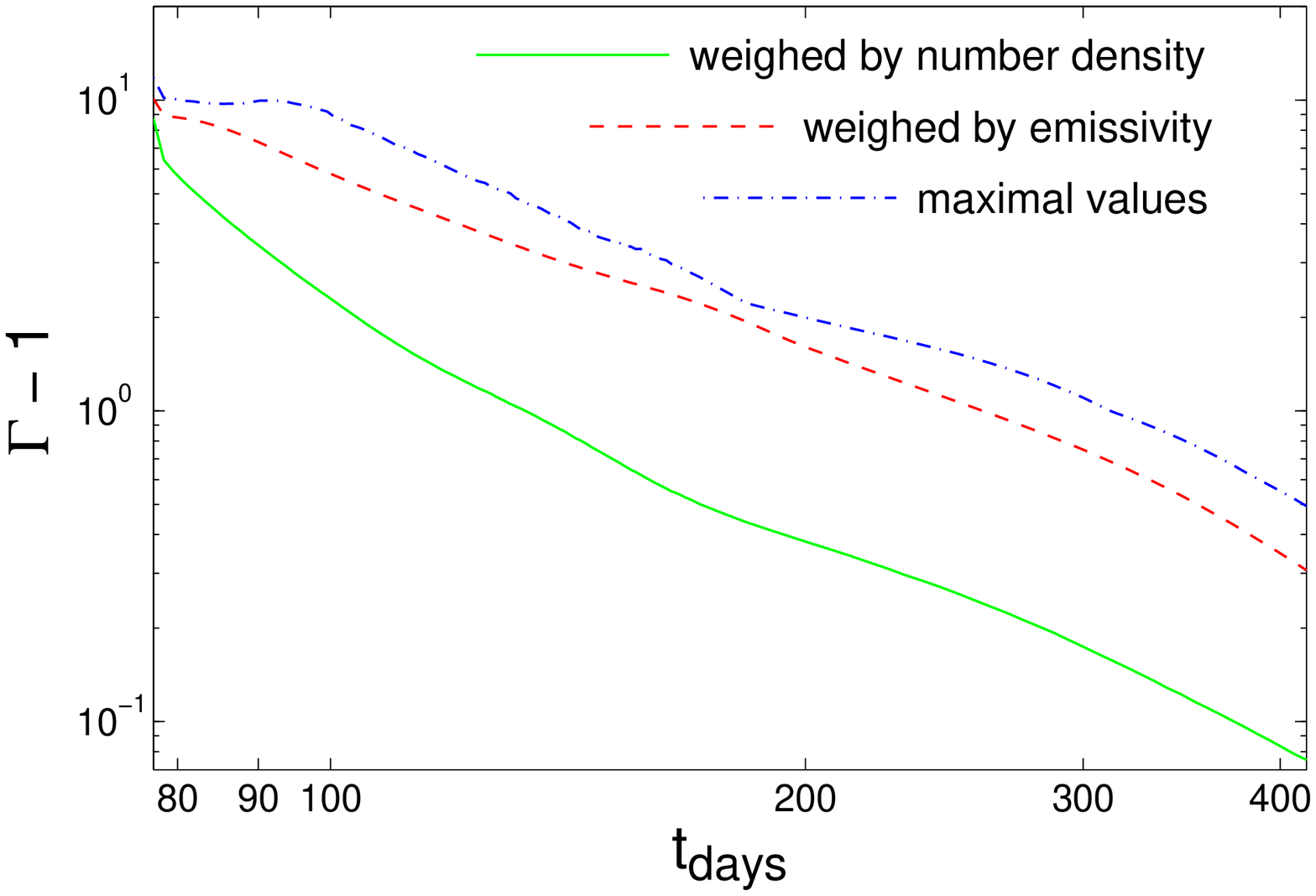}
\includegraphics[width=3.97cm]{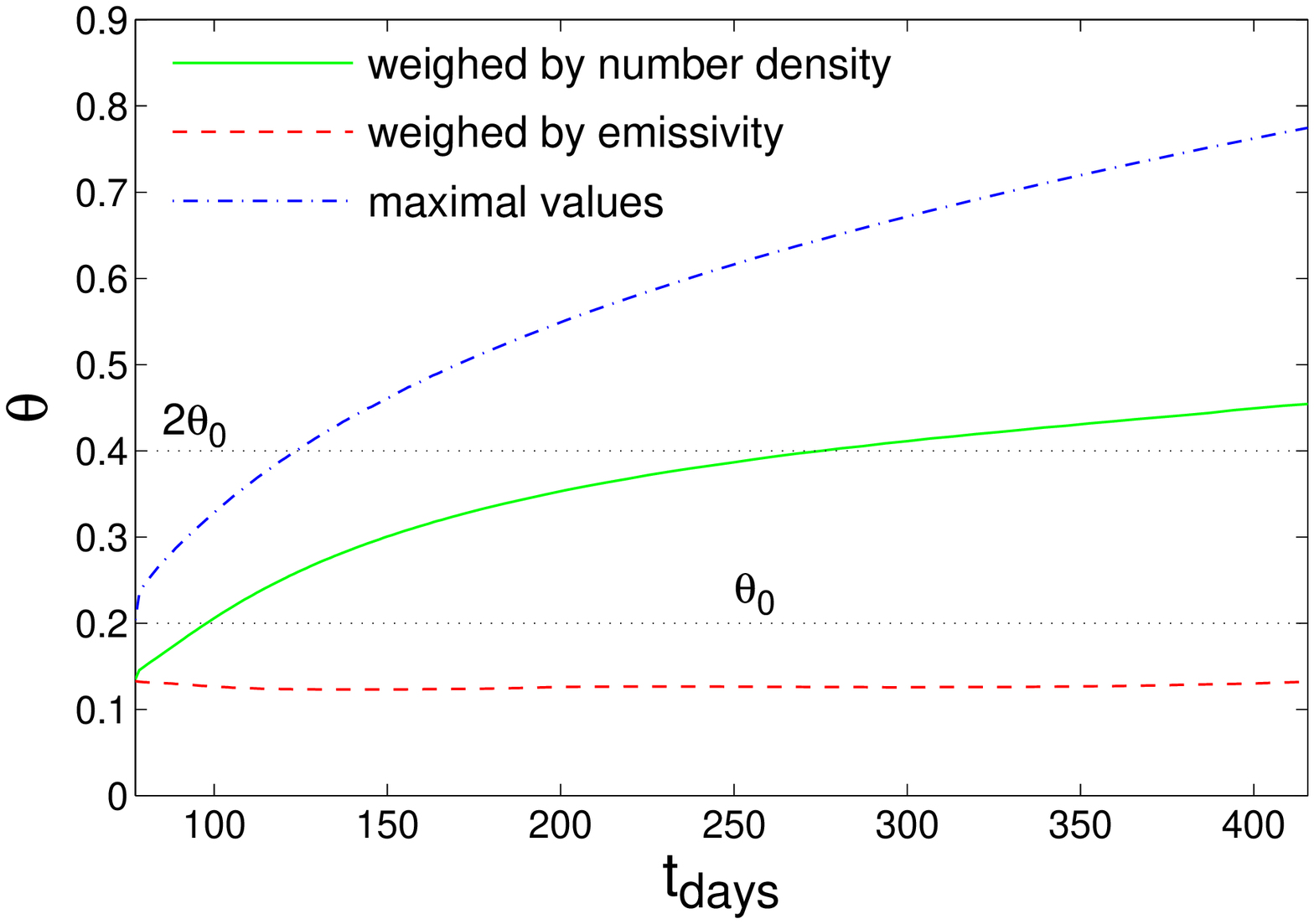}
\end{center}
\caption[]{The radius $R$ ({\it left frame}), Lorentz factor $\Gamma-1$
  ({\it middle frame}) and opening angle $\theta$ of the jet ({\it
    right frame}), as a function of the lab frame time in days \cite{Granot01}.}
\label{averages}
\end{figure}

However, recent 2D hydrodynamic simulations \cite{Granot01} show
that these assumptions are not a good approximation of a
realistic jet. Figure \ref{3Djet} shows the jet at the last time
step of the simulation. The matter at the sides of the jet is
propagating sideways (rather than in the radial direction) and is
slower and much less luminous compared to the front of the jet.
The shock front is egg-shaped, and quite far from being
spherical. Figure \ref{averages} shows the radius $R$, Lorentz
factor $\Gamma$, and opening angle $\theta$ of the jet, as a
function of the lab frame time. The rate of increase of $\theta$
with $R\approx ct_{\rm lab}$, is much lower than the exponential
behavior predicted by simple models \cite{Rhoads99}. The value of
$\theta$ averaged over the emissivity is practically constant,
and most of the radiation is emitted within the initial opening
angle of the jet. The radius $R$ weighed over the emissivity is
very close to the maximal value of $R$ within the jet, indicating
that most of the emission originates at the front of the
jet\footnote{This implies that the expected rate of orphan optical
afterglows should be smaller than estimated assuming significant
sideways expansion!}, where the radius is largest, while $R$
averaged over the density is significantly lower, indicating that
a large fraction of the shocked matter resides at the sides of
the jet, where the radius is smaller. The Lorentz factor $\Gamma$
averaged over the emissivity is close to its maximal value,
(again since most of the emission occurs near the jet axis where
$\Gamma$ is the largest) while $\Gamma$ averaged over the density
is significantly lower, since the matter at the sides of the jet
has a much lower $\Gamma$ than at the front of the jet. The large
differences between the assumptions of simple dynamical models of
a jet and the results of 2D simulations, suggest that great care
should be taken when using these models for predicting the light
curves of jetted afterglows. Since the light curves depend
strongly on the hydrodynamics of the jet, it is very important to
use a realistic hydrodynamic model when calculating the light
curves.

Granot et al. \cite{Granot01} used 2D numerical simulations of a jet running
into a constant density medium to calculate the resulting light
curves, taking into account the emission from the volume of the
shocked fluid with the appropriate time delay in the arrival of
photons to different observers. They obtained an achromatic jet break
for $\nu>\nu_m(t_{\rm jet})$ (which typically includes the optical and
near IR), while at lower frequencies (which typically include the
radio) there is a more moderate and gradual increase in
the temporal index $\alpha$ at $t_{\rm jet}$, and a much more prominent
steepening in the light curve at a latter time when $\nu_m$ sweeps
past the observed frequency. The jet break appears sharper and occurs
at a slightly earlier time for an observer along the jet axis,
compared to an observer off the jet axis (but within the initial
opening angle of the jet). The value of $\alpha$
after the jet break, for $\nu>\nu_m$, is found to be slightly larger than $p$
($\alpha=2.85$ for $p=2.5$).

Somewhat surprisingly we find that in spite of the different
hydrodynamic behavior the numerical simulations show a jet break
at roughly the same time as the analytic estimates. This
encourages us to trust the current estimates of the jet opening
angles. However, we should search for an intuitive explanation
for the nature of the hydrodynamic behavior and for a simple
analytic model that would predict it.


\begin{thebibliography}{8.}
\addcontentsline{toc}{section}{References}

\bibitem{BM76}R.D. Blandford,  C.F. McKee:
 Phys. of Fluids, {\bf 19}, 1130 (1976).

\bibitem{CPS98} E. Cohen, T. Piran, R. Sari: Ap. J., \textbf{509}, 717 (1998).

\bibitem{CP99} E. Cohen, T. Piran: Ap. J., \textbf{518}, 346 (1999).

\bibitem{Usov94}V.V. Usov: MNRAS, {\bf 267},1035 (1994)

\bibitem{SPN98} R. Sari, T. Piran, R. Narayan: Ap. J. Lett., \textbf{497}, L17 (1998).

\bibitem{Sari97} R. Sari: Ap. J. Lett., \textbf{489}, L37 (1997).

\bibitem{SP97} R. Sari, T. Piran: MNRAS, \textbf{287}, 110 (1997).

\bibitem{GPS99} J. Granot, T. Piran, R. Sari: Ap. J., \textbf{513}, 679 (1999).

\bibitem{CheLi99} R.A. Chevalier, Z.-Y. Li: Ap. J. Lett., \textbf{520}, L29 (1999).

\bibitem{SariEsin01} R. Sari, A.A. Esin: Ap. J., \textbf{548}, 787 (2001).

\bibitem{Pi94} T. Piran: in AIP Conference Proceedings {\bf 307},
{\it Gamma-Ray Bursts, Second Workshop, Huntsville, Alabama,
1993}, Fishman, G.J., Brainerd, J.J., \& Hurley, K., Eds., (New
York: AIP), p. 495. (1994)

\bibitem{Rhoads99} J.E. Rhoads: Ap. J., \textbf{525}, 737 (1999)

\bibitem{Pi00} T. Piran: Phys. Rep., \textbf{333}, 529 (2000)

\bibitem{KP00} P. Kumar \& A. Panaitescu: Ap. J., \textbf{541}, L9 (2000)

\bibitem{SPH99} R. Sari, T. Piran, T. Halpern: Ap. J., \textbf{519}, L17 (1999).

\bibitem{PM99} A. Panaitescu \& P. M\'esz\'aros: Ap. J., \textbf{526},
  707 (1999)

\bibitem{MSB00} R. Moderski, M. Sikora, T Bulik: Ap. J., \textbf{529}, 151 (2000)

\bibitem{HDL00} Y. Huang, Z. Dai \& T. Lu: A\&A \textbf{355}, L43 (2000)

\bibitem{Granot01} J. Granot, et al.: These proceedings (astro-ph/0103038) (2001)




\end{thebibliography}
\end{document}